\documentclass[doublecol]{modepl} 
\usepackage{graphicx}
\title{Bound-free pair production cross section in heavy-ion colliders\\
from the equivalent photon approach}
\shorttitle{Bound-free pair production}

\author{A. Aste\inst{1}}
\shortauthor{A. Aste}

\institute{                    
  \inst{1} Department of Physics, University of Basel,
Klingelbergstrasse 82, 4056 Basel, Switzerland}


\abstract{Exact calculations of the electron-positron pair production
by a single photon in the Coulomb field of a nucleus with simultaneous
capture of  the electron into the K-shell are discussed for different
nuclear charges.
Using the equivalent photon method of Weizs\"acker and 
Williams, a simple expression for the bound-free production 
of $e^{+} e^{-}$ pairs by colliding very-high-energy fully stripped 
heavy ions is derived for nuclei of arbitrary charge. 
}

\begin{document}

\maketitle

\section{Introduction}
There has been some theoretical and experimental interest in the process of 
single-photon pair production during the last few decades. Precise knowledge of the
single-photon bound-free pair production cross section in the electrostatic
field of a fully stripped nucleus allows to calculate
the bound-free pair production at the interaction point of a heavy ion
collider by using the equivalent photon method of Weizs\"acker and Williams \cite{Weiz,Willi,Baur}.
Recently, first observations of beam losses have been reported
from measurements performed at the BNL Relativistic Heavy Ion Collider (RHIC),
where a beam of $^{63}\mbox{Cu}^{29+}$ ions with 100 GeV/nucleon has been used \cite{Bruce}.
In this paper, we derive a simple and compact expression for the  
bound-free pair production in relativistic heavy-ion collisions, which is expected
to be a major luminosity limit for the Large Hadron Collider (LHC) when it operates with
heavy ions because the localized energy deposition by the lost ions which
have captured an electron may quench superconducting magnet coils \cite{Klein}.

\section{The equivalent photon method}
The simplest way to  get a reliable estimate of the cross section in RHI collisions
is provided by the equivalent photon method, which is originally due to Fermi 
\cite{Fermi}, and later on developed by Weizs\"acker and Williams.
Within this framework, the target ion is considered as fixed.
The projectile with impact parameter $b$ is assumed to move on a straight line with
velocity $v$ and a relativistic Lorentz factor $\gamma_p=2 \gamma_c^2-1$,
where $\gamma_c$ is the Lorentz boost of the ions in the center-of-mass frame.
The projectile is accompanied by its contracted electromagnetic field,
which corresponds to a spectrum of equivalent photons, given by
\begin{equation}
N(\omega, b) = \frac{Z_{p}^{2} \alpha}{\pi^2} \biggl( \frac{\omega^2} {\gamma_{p}^{2} v^4} \biggr)
\biggl[ K_{1}^{2}(x) + \frac{1}{\gamma_{p}^{2}} K_{0}^{2}(x) \biggr],
\end{equation}
where $K_{0}$($K_{1}$) are the modified Bessel functions of the second kind of order zero (one), $x=\omega b/ \gamma_p v$, and $\alpha$ is the
fine-structure constant. The cross section for an 
electromagnetic process in a highly relativistic collision is then
obtained by integrating the single-quantum cross section over the 
frequency spectrum and from a minimum impact parameter $b_{min}=R$, 
which was, in our case, chosen to be the Compton wavelength as the typical
length scale for pair production, to infinity:
\begin{equation}
n(\omega) = \int \limits_{R}^{\infty} 2 \pi b N(\omega,b) db  ,
\end{equation}
\begin{equation}
\sigma = \int n(\omega) \sigma_{\gamma} (\omega) \frac{d\omega}{\omega},
\end{equation}
where $\sigma$ is now the total cross section of the electromagnetic
process.
The integration of $N(\omega,b)$ over $b$ can be carried out
to give \cite{BeB}
\begin{eqnarray*}
n( \omega ) = 2 \pi \int \limits_{R}^{\infty} b N(\omega,b) db
\end{eqnarray*}
\begin{equation}
= \frac{2}{\pi} \frac{Z_{p}^{2} \alpha}{v^2}
\biggl[\zeta K_0(\zeta) K_1(\zeta) - \frac{v^2 \zeta^2}{2}
(K_{1}^{2}(\zeta) -K_{0}^{2}(\zeta)) \biggr],
\end{equation}
where $\zeta = \omega R/\gamma_p v$ is an adiabatic cutoff parameter.
For $\gamma \gg 1 $ (except for extreme low-energy frequencies, 
satisfying the relationship $\omega R/c \ll 1$), 
one can use the excellent approximation
\begin{equation}
n(\omega) = \frac{1}{\pi} Z_{p}^{2} \alpha
\log \biggl[ \biggl( \frac{\delta}{\zeta} \biggl)^2 +1 \biggl]
\cong \frac{2}{\pi} Z_{p}^{2} \alpha \log \biggl( \frac{\delta}
{\zeta} \biggl)  , \label{nphot}
\end{equation}
where $\delta = 0.68108...$ is a number related to Euler's constant.
In the limit of very large frequencies, $\omega \gg \gamma_p v/R$, an
adiabatic cutoff sets in and one has
$n(\omega) \cong (\alpha Z_p^2/2) e^{-2 \omega R/\gamma_p v}$.
For practical calculations, it is usually sufficient to
set the velocity $v$ equal to the speed of light $c$.

Calculations of the single-photon bound-free pair production
cross section $\sigma_{e^+e^-}^K$ with simultaneous capture
of the electron into the K-shell of a target nucleus with
nuclear charge number $Z_t$ have been presented in \cite{Aste}.
In the meantime, the program used in \cite{Aste} has been refined
to higher precision, leading to small corrections of the originally calculated K-shell cross sections of some few percent.
In \cite{Helmar}, it has been shown that
the higher shells contribute about $20\%$ of the K-shell capture to
the total bound-free pair production cross section $\sigma_{e^+e^-}$,
i.e., $\sigma_{e^+e^-}=\zeta(3) \sigma_{e^+e^-}^K$ provides a good
approximation for $\sigma_{e^+e^-}$, where $\zeta$ is the
Riemann zeta function defined by
\begin{equation}
\zeta(x)=\sum \limits_{n=1}^{\infty} \frac{1}{n^x}
\end{equation}
for $x>1$, and $\zeta(3) \simeq 1.2020569$.

\section{Single-photon cross sections}

\begin{figure}
\begin{center}
    \mbox{\includegraphics[width=3.10in]{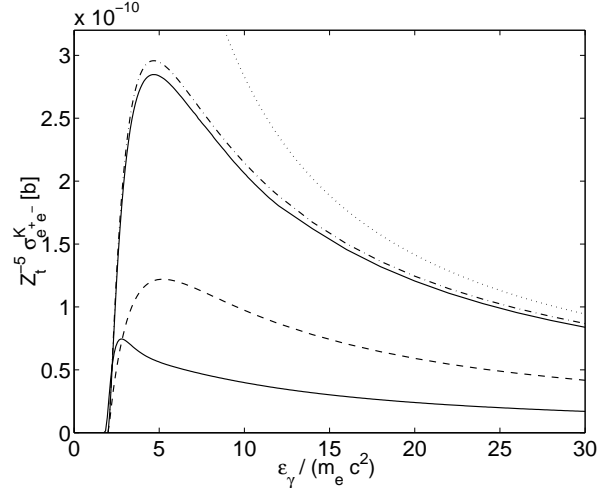}}
        \caption{Single-photon cross section for
          pair production with K-shell capture as a function of the
          photon energy $\epsilon_\gamma=\hbar \omega$ for 
          different nuclear charges $Z_t$. The cross sections
          are scaled with $Z_t^5$. The dash-dotted curve displays the
          Sauter cross section according to eq. (\ref{Sauter_full}),
          which agrees well with the numerical result for $Z_t=1$
          (upper solid curve).
          The dashed curve corresponds to $Z_t=29$ (copper), the lower
          solid curve to $Z_t=92$ (uranium). The dotted curve shows
          the high-energy limit of the Sauter cross section
          according to eq. (\ref{Saut}).}
        \label{fig1}
\end{center}
\end{figure}

Theoretical calculations using semirelativistic Som\-mer\-feld-Maue wave 
functions for the positron and a semi-relativistic approximation for the bound electron wave functions valid for $\alpha Z_t \ll 1$
show that, for $\hbar \omega \gg 1$ and light nuclei, the cross 
section $\sigma_{e^+e^-}^K$ is approximately given by the
high-energy limit of the so-called Sauter cross section \cite{Sauter}
\begin{equation}
\sigma_{e^+ e^-}^{K} = 4 \pi \lambda_c^2 \alpha^6 Z_t^5 \frac{m_e c^2}{\hbar \omega}, \label{Saut}
\end{equation}
where $m_e$ is the electron mass and $\lambda_c$ the Compton wavelength.

The high-energy limit of the Sauter cross section
overestimates the exact cross sections at low photon energies.
Indeed, Sauter originally derived an exact result in the limit
of vanishing $Z_t$ for the K-shell photoelectric effect, which can
be transcribed into the corresponding cross section for pair creation
\cite{Sorensen} ($m_e=\hbar=c=1$)
\begin{eqnarray*}
\sigma_{e^+ e^-}^{K (Sauter)}= 4 \pi \lambda_c^2  \alpha^6 Z_t^5
\frac{{p_+}^3}{\omega^5}
\end{eqnarray*}
\begin{equation}
\times \Biggl(\frac{E_+ (E_+ +2)}{E_+ -1} - \frac{4}{3}
-\frac{E_+ +2}{2 p_+ (E_+ -1)} \log \Biggl[ \frac{E_+ + p}{E_+ -p} \Biggl]
\Biggl), \label{Sauter_full}
\end{equation}
where $E_+$ and $p_+=(E_+^2-1)^{1/2}$ is the energy and the momentum of the
positron, respectively.
The photon threshold energy for pair production
$\omega_{th}=1+\gamma_e$ is given by the sum of the positron rest
energy and the energy $\gamma_e=(1-\alpha^2 Z_t^2)^{1/2}$
of the bound electron in the K-shell.
Generally, the positron energy is given by $E_+ = \omega-\gamma_e$,
and in the limit $Z_t \rightarrow 0$ one obviously has $\omega=E_+ +1$.

A comparison of the Sauter cross section and exact
numerical calculations for $Z_t=1,29$, and 92, is shown
in Fig. \ref{fig1}.
The numerical result for $Z_t=1$ agrees well with eq. (\ref{Sauter_full}).
However, at high energies, the cross section for hydrogen is smaller
by a factor of $0.971$ than the Sauter cross section.
This effect becomes more pronounced for target nuclei  with large $Z_t$,
such that eq. (\ref{Saut}) must be modified by an additional factor
$f(Z_t)$ according to
\begin{equation} 
\sigma_{e^+ e^-}^{K} = 4 \pi \lambda_c^2 \alpha^6 Z_t^5 f(Z_t) \frac{1}{\omega} .
\label{cs}
\end{equation}
Some values for $[Z_t,f(Z_t)]$ obtained from exact numerical calculations
are $[92,0.196]$, $[82,0.216]$, $[29,0.484]$ and [1,0.971].
These values agree very well with those found by Pratt \cite{Pratt} and
coincide with the values calculated by Agger and Sorensen \cite{Agger}.

A useful approximation of $f(Z_t)$ is given by
\begin{equation}
f(Z_t)=\biggl( \frac{53}{100} + \frac{2 \alpha Z_t}{9}
\biggr) ^{2 \pi \alpha Z_t}, \label{Sauterfit}
\end{equation}
which reproduces $f(Z_t)$ to an accuracy of $2 \%$ in the physically
relevant range $0 \leq Z_t \leq 92$.

As a further effect one observes that the position and the shape
of the cross section peak varies for different nuclear charges.
Whereas the peak is shifted to higher photon energies for light
and medium-heavy nuclei, it is located at lower energies with
respect to the Sauter cross section peak for $Z_t \ge 62$.

\section{Pair production in heavy-ion collisions}
Eq. (\ref{Sauter_full}) in conjunction with the equivalent photon method
allows to derive a precise numerical expression for the total
bound-free pair production in relativistic heavy-ion collisions
in the case of small target charge $\alpha Z_t \ll 1$
(with $R=1$ and $v=1$ in eq. (\ref{nphot})), given by
\begin{eqnarray*}
\sigma_{RHI} = \zeta(3) \sigma_{RHI}^K =
\zeta(3) \int \limits_{2}^\infty n(\omega) \sigma_{e^+ e^-}^{K (Sauter)}(\omega)
\frac{d \omega}{\omega}
\end{eqnarray*}
\begin{eqnarray*}
\simeq 8 \zeta(3) \lambda_c^2 \alpha^7 Z_t^5 Z_p^2
\int \limits_{2}^{\infty}
\log \biggl( \frac{\delta \gamma_p}{\omega} \biggr)
\sigma_{e^+ e^-}^{K (Sauter)}(\omega)
\frac{d \omega}{\omega}
\end{eqnarray*}
\begin{equation}
=8 \zeta(3) \lambda_c^2 \alpha^7 Z_t^5 Z_p^2
\biggl[ c_1 \log \gamma_p - c_2 \biggr],
\label{rhic1}
\end{equation}
with $c_1=0.21746$ and $c_2=0.58706$.
Eq. (\ref{rhic1}) still overestimates the cross section, since the single-photon cross section is suppressed by a factor $f(Z_t)$
at large photon energies. Additionally, the shape of the single
photon cross section peak has some impact on $\sigma_{RHI}$,
since the photon flux generated by the Coulomb field of the
projectile becomes large at low equivalent photon energies.
The defect can be remedied by a minor modification of eq. (\ref{rhic1})
\begin{equation}
\sigma_{RHI} =8 \zeta(3) \lambda_c^2 \alpha^7 Z_t^5 Z_p^2 f(Z_t)
\biggl[ c_1(Z_t) \log \gamma_p - c_2(Z_t) \biggr],
\label{rhic2}
\end{equation}
where a numerical fit for $0 \leq Z_t \leq 92$ gives
\begin{equation}
c_1(Z_t) \cong \Biggl[ 1- \frac{Z_t}{2.5 \cdot 10^2} +
\frac{Z_t^2}{1.5 \cdot 10^4}-\frac{Z_t^3}{7.0 \cdot 10^6} \Biggr] c_1,
\label{coeff1}
\end{equation}
\begin{equation}
c_2(Z_t) \cong \Biggl[ 1- \frac{Z_t}{2.0 \cdot 10^2} +
\frac{Z_t^2}{1.2 \cdot 10^4}-\frac{Z_t^3}{7.0 \cdot 10^6} \Biggr] c_2,
\label{coeff2}
\end{equation}
with $c_{1,2}=c_{1,2}(0)$ given above.

Expressing the cross sections directly in barn leads to
\begin{equation}
\sigma_{RHI}=1.58 \cdot 10^{-11} Z_t^5 Z_p^2 f(Z_t)
[c_1(Z_t) \log(\gamma_p)-c_2(Z_t)]. \label{handyformula}
\end{equation}

The formula above describes the equivalent
photon cross section with an accuracy better than $2\%$ for
$Z_t \leq 92$ and typical RHIC and LHC energies.
It should be mentioned that it has already been suggested
in \cite{Baltz} that the cross section $\sigma_{RHI}$
can be established in the form
\begin{equation}
\sigma_{RHI}=A \log \gamma_p +B. \label{logform}
\end{equation}
The coefficients $A$ and $B$ were obtained from numerical calculations. In our case, the coefficients
follow directly from eqns. (\ref{coeff1},\ref{coeff2},\ref{handyformula}). 

Cross sections for several nuclei and energies have been calculated, e.g.,
by Meier {\it{et al.}} \cite{Helmar} and Baltz {\it{et al.}} \cite{Baltz2}.
For a RHIC Cu-Cu collision with 100 GeV/nucleon, one obtains with
$\gamma_c \simeq 107$ and $\gamma_p =2 \gamma_c^2-1$ a cross section
of $0.198$ barn, in agreement with recent experimental measurements
\cite{Bruce}. In the case of the LHC, one obtains from
eq. (\ref{handyformula}) for a Pb-Pb collision
with $\gamma_c=2957$ a cross section of $220$ barn for K-shell capture,
compared to $225$ barn in \cite{Helmar}.
For $Z_p=Z_t=20$ and $\gamma_c=125$ ($\gamma_c=3750$),
the cross section for K-shell capture obtained from the
equivalent photon method is
$1.57 \cdot 10^{-2}$ barn ($2.96 \cdot 10^{-2}$ barn).
The corresponding values found in \cite{Helmar} are
$1.61 \cdot 10^{-2}$ barn and $2.92 \cdot 10^{-2}$ barn, respectively.
A cross section of 89 barn for Au-Au collisions at RHIC energy
given in \cite{Baltz2} is also in excellent agreement with
eq. (\ref{handyformula}).
This demonstrates the validity of the equivalent photon method for
the full physical range of charge numbers $Z$ and large $\gamma_c$.

\begin{figure}
\begin{center}
    \mbox{\includegraphics[width=3.10in]{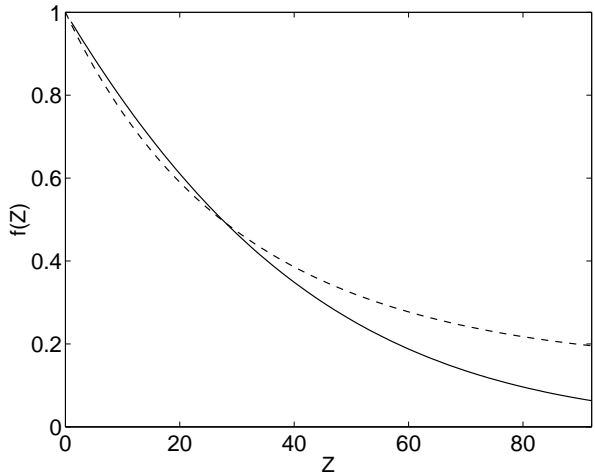}}
        \caption{Comparison of the $Z$-dependence of
         $\sigma_{e^+ e^-}^K$, expressed by $\tilde{f}(Z)$ from
         eq. (\ref{ftilde}) (solid line), and exact values for
         $f(Z)$ (dashed line).}
        \label{fig2}
\end{center}
\end{figure}

As an interesting final remark, we briefly compare the results
obtained above to approaches presented in \cite{BeB} and
in \cite{Dolci}. Based on Sommerfeld-Maue wave functions for
positrons and semi-relativistic electron wave functions
constructed from the non-relativistic electron wave functions
by including perturbative correction terms of the order $\alpha Z$,
an approximate expression for pair production with K-shell capture
was derived in \cite{BeB}
($Z=Z_t=Z_p$)
\begin{equation}
\sigma_{RHI}^K = \frac{33}{20} \lambda_c^2 \alpha^7 Z^7
\frac{2 \pi \alpha Z}{e^{2 \pi \alpha Z}-1} \Bigl[
\log ( \delta \gamma_p/2)-5/3 \Bigr].
\end{equation}
The numerical factor $33/20=1.65$ agrees in a satisfactory manner
with the corresponding factor $8c_1 \cong 1.74$ appearing in
eq. (\ref{rhic1}). The deviation of the cross section from the
$Z^5$-scaling is basically described by
\begin{equation}
\tilde{f}(Z)=\frac{2 \pi \alpha Z}{e^{2 \pi \alpha Z}-1}.
\label{ftilde}
\end{equation}
A comparison of $\tilde{f}(Z)$ with $f(Z)$ obtained from
exact numerical calculations is presented in Fig. \ref{fig2}.
Whereas $\tilde{f}(Z)$ is very acceptable for $\alpha Z \ll 1$,
it leads to an underestimation of cross sections for
heavy nuclei by a factor of about 2.
The approach presented in \cite{Dolci} uses the same bound
electron wave functions as in \cite{BeB}, however, positron wave
functions are described by normalized plane waves in conjunction with
a correction term proportional to $\alpha Z$.
The approximation in \cite{Dolci} suffers from the opposite effect,
i.e., cross sections for nuclei with large charge numbers
turn out to be too large by factors exceeding 2. This observation
shows that is it crucial to work with exact solutions of the Dirac
equation in order to get reliable results in the present case.

\section{Conclusions}
A heuristic analytic fit for the characteristic function $f(Z)$
describing the single-photon pair production with capture on
fully stripped nuclei at high energies is presented in
eq. (\ref{Sauterfit}). This fit allows to describe the single-photon
pair production cross section at high energies with an accuracy
comparable to presently available numerical calculations.
Furthermore, eq. (\ref{handyformula}) in conjunction with eqns.
(\ref{Sauterfit},\ref{coeff1},\ref{coeff2}) allows to evaluate  
the bound-free pair production cross section in relativistic
heavy-ion collisions and the corresponding coefficients $A$ and
$B$ in eq. (\ref{logform}) in a simple and reliable manner.
It should be mentioned that in currently available
calculations, multi-photon interactions between the photon-emitting
nuclei are neglected, which might lead to an additional correction of
the cross section of a few percent, probably with different energy
dependence from the overall cross-section.

\end{document}